\documentclass{article}
\usepackage[pdftex]{graphics}
\usepackage{amssymb}

\begin{document}

\title{Solving the Parity Problem with Rule 60 in Array Size of the Power of Two}

\author{Shigeru Ninagawa \\
Kanazawa Institute of Technology,Japan \\
University of the West of England, United Kingdom}

\maketitle

\begin{abstract}
In the parity problem, a given cellular automaton has to classify any initial configuration into two classes according to its parity. Elementary cellular automaton rule 60 can solve the parity problem in periodic boundary conditions with array size of the power of two. The spectral analysis of the configurations of rule 60 at each time step in the evolution reveals that spatial periodicity emerges as the evolution proceeds and the patterns with longer period split into the ones with shorter period. This phenomenon is analogous to the cascade process in which large scale eddies split into smaller ones in turbulence. By measuring the Lempel-Ziv complexity of configuration, we found the stepping decrease of the complexity during the evolution. This result might imply that a decision problem solving process is accompanied with the decline of complexity of configuration.
\end{abstract}

\section{Introduction}
A wide variety of cellular automata (CAs) have been invented to perform various computational tasks. In the parity problem (PP), CA must classify any initial configuration into two classes according to its parity. Specifically, the initial configuration $\sigma(0)$ must be evolved to the configuration where the state of every cell is equal to $P(\sigma(0)) \in \{0, 1 \}$, where $P(\sigma(0))$ denotes the parity of $\sigma(0)$. Several methods to tackle the PP have been proposed such as using nonuniform CAs~\cite{Sipper} or applying several one-dimensional and two-state, three-neighbor CA (called elementary CA, ECA) rules in succession~\cite{Lee, Martins}. Tsuchiya and Komito suggested that ECA rule 60 can solve the PP if array size $N$ is restricted to $2^n$ ($n$ is positive integer) by performing extensive computer experiments~\cite{Tsuchiya,Komito}. Thereafter a mathematical proof was given~\cite{Nina} using the characteristic polynomial~\cite{Martin}. In this article we study the process of solving the PP by ECA rule 60 from the two viewpoints, namely, the transition of dynamical systems and the computation of decision problem. This article is organized as follows. In section 2, we explain the formulation of solving the PP by additive cellular automata. Next to study the solving process of the PP by ECA rule 60, we use spectral analysis in section 3 and Lempel-Ziv complexity in section 4. Finally we discuss the meaning of the results especially connected with the solving process of decision problem in section 5.

\begin{figure}
\begin{center}
\scalebox{0.55}{\includegraphics{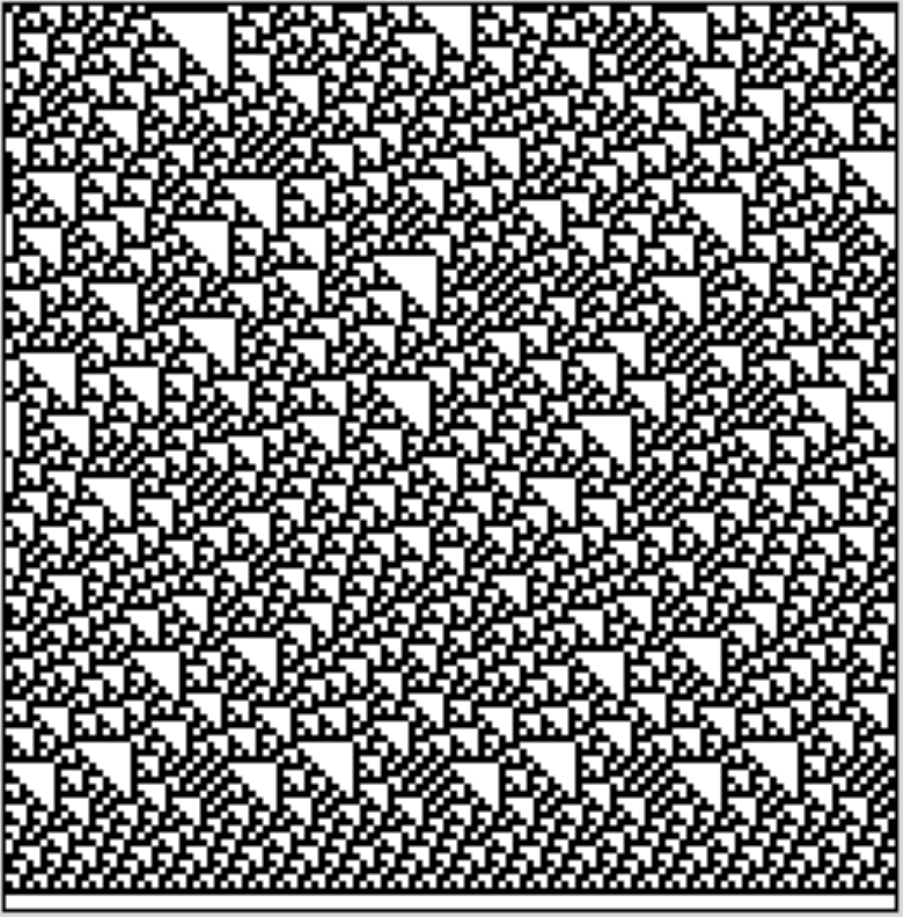}}
 \scalebox{0.55}{\includegraphics{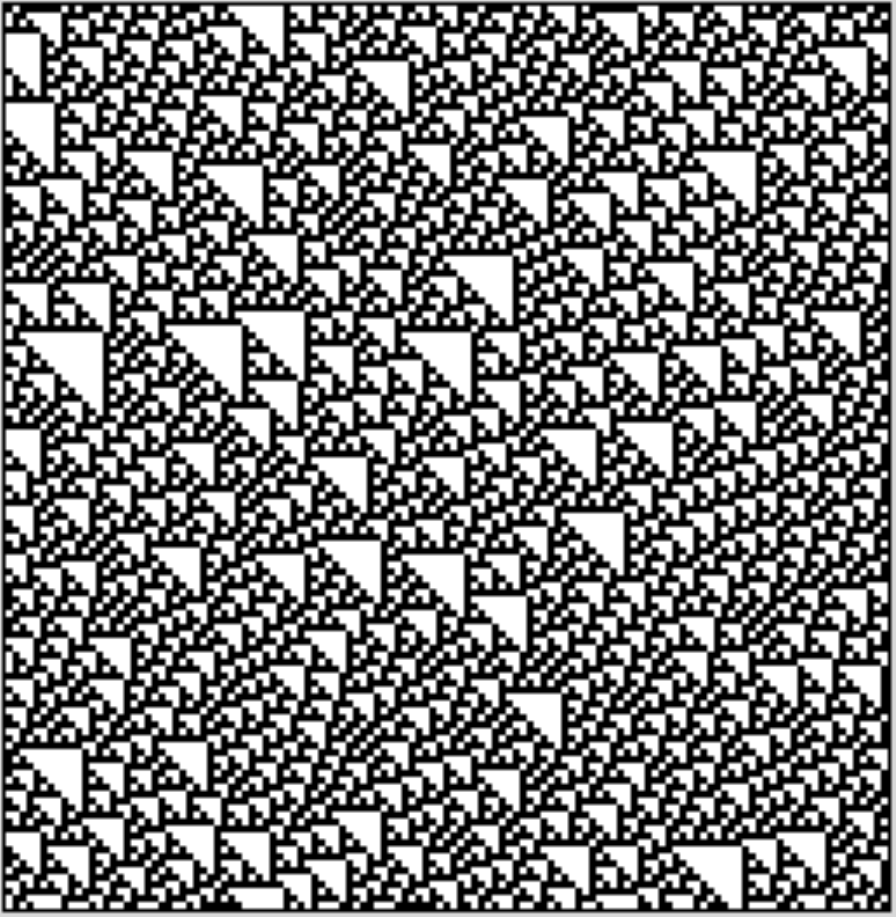}}
\end{center}
\caption{Space-time patterns of rule 60 starting from a random initial configuration. Array size is
128 (left) and 127 (right).}
\label{fig:r60conf}
\end{figure}

\section{Additivity}
Let $\phi$ denote the global transition function of a CA with binary state $\{0, 1\}$. The configuration $\sigma(t+1)$ at time step $t+1$ is given by the mapping $\phi[\sigma(t)] = \sigma(t+1)$. Additivity imposes the following property on the global transition function $\phi$
\begin{equation}
\phi[\rho \oplus \tau ] = \phi[\rho] \oplus \phi[\tau],
\label{eq:add}
\end{equation}
for any two configurations $\rho$ and $\tau$ and, where $\oplus$ denotes the sum modulo two of the state of each cell. Throughout this article the array size $N$ is limited to $2^n$ ($n$ is positive integer) and periodic boundary conditions are employed. 

\begin{figure}
\begin{center}
\scalebox{0.7}{\includegraphics{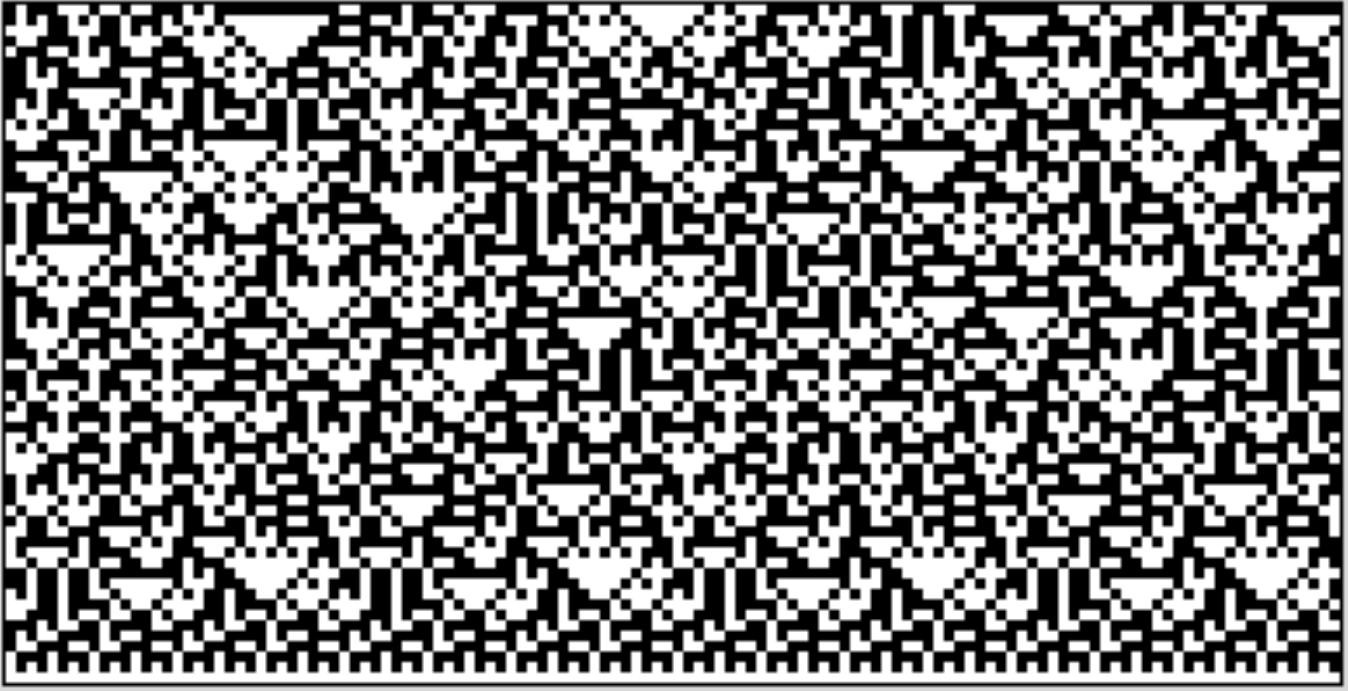}}
 \scalebox{0.7}{\includegraphics{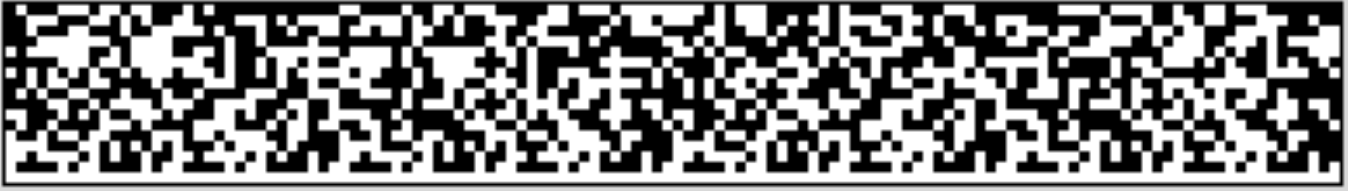}}
\end{center}
\caption{Space-time patterns of rule 90 (top) and of the CA rule defined
 in Eq.~(\ref{eq:Gentrns}) with $r$ = 8 (bottom) starting from a random
 initial configuration with odd parity. Array size is 128.}
\label{fig:r90conf}
\end{figure}

Let us consider CA rules described as follows.
\begin{equation}
a_i(t+1) = a_{i-r}(t) + a_{i}(t),
\label{eq:Gentrns}
\end{equation}
where $a_{i}(t)$ denotes the state of the $i$-th cell at time step t that is represented by integers modulo 2 and $r$ is positive integer. The transition rule defined by Eq.~(\ref{eq:Gentrns}) satisfies the additivity (Appendix A). If $r$ is odd, the state of any one cell at $t = 2^n-1$ is equal to the parity of initial configuration. If $r$ is even, the parity of any consecutive $2^k$ cells at $t = 2^{n-k} - 1$ is equal to the parity of initial configuration where $k$ is the highest power of 2 that divides $r$ (Appendix B). In both cases, the evolution eventually reaches the null configuration in which all cells have state zero. In other words, the null configuration is a fixed point of configuration space that attracts evolution.

Figure~\ref{fig:r60conf} shows the space-time patterns of the evolution of rule 60 starting from a random initial configuration of array size equal to $2^7=128$ (left) and 127 (right). White squares represent cells in state zero and black squares represent cells in state one. Time goes from top to bottom. As the parity of the initial configuration on the left of Fig.~\ref{fig:r60conf} is odd, every cell has state one at $t$ = 127. The configurations seem to turn into periodic as the times step comes close to $t$ = 127 while the evolution on the right does not reach stable configuration.

Figure~\ref{fig:r90conf} shows the space-time patterns of rule 90 (top) and the CA rule defined in Eq.~ (\ref{eq:Gentrns}) with $r$ = 8 (bottom) starting from the same configuration as in Fig.~\ref{fig:r60conf}. The rule obtained by letting $r$ = 2 in Eq.~(\ref{eq:Gentrns}) is virtually the same as that of rule 90 because they differ only by a translation. In the case of rule 90, the configuration becomes the repetition of Ž¡ŽÆ01Ž¡ŽÇ at time step $t$ = 63. In the case of the rule with $r$ = 8, the configuration becomes the repetition of Ž¡ŽÆ01111010Ž¡ŽÇ at time step $t$ = 15. In both cases,  the result means the parity of the initial configuration is odd.

From now on we focus our attention on ECA rule 60 that is obtained by setting $r$ = 1 in Eq.~ (\ref{eq:Gentrns}). Throughout this paper we exclusively use initial configurations with odd parity, because those can occur only as initial configurations (Appendix C). The evolution starting from a configuration with odd parity takes longest time steps to reach the null configuration. On the other hand, it is possible that the evolution with an initial configuration with even parity reaches the null configuration before the time step $t = 2^n-1$.
The state transtion graph of rule 60 with array size $N = 1 \sim 15$ is presented in \cite{Wuensche}.

\begin{figure}
\begin{center}
\scalebox{0.84}{\includegraphics{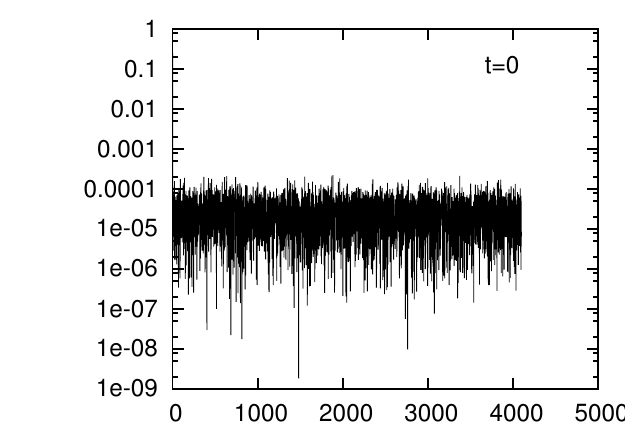}}
\scalebox{0.84}{\includegraphics{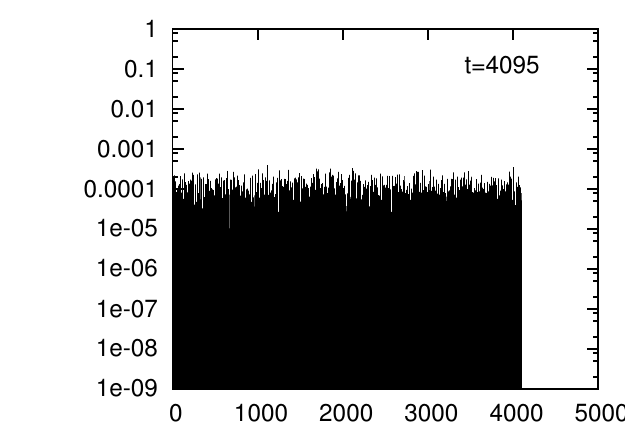}}
\scalebox{0.84}{\includegraphics{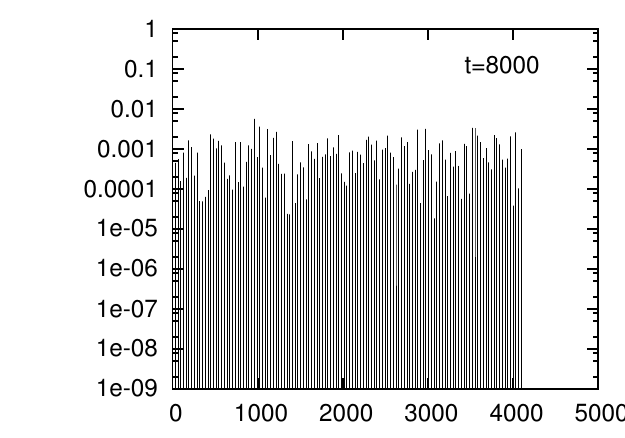}}
\scalebox{0.84}{\includegraphics{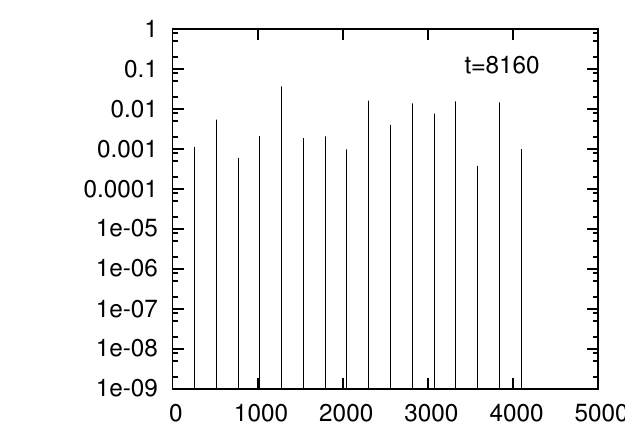}}
\scalebox{0.84}{\includegraphics{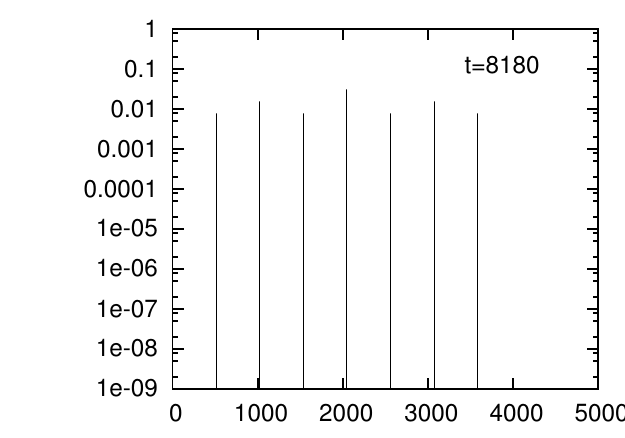}}
\scalebox{0.84}{\includegraphics{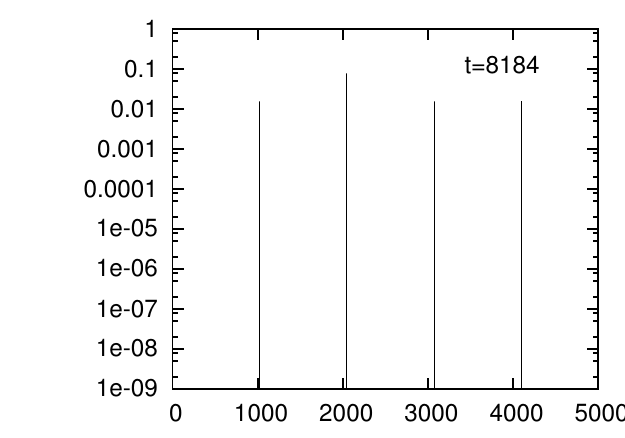}}
\scalebox{0.84}{\includegraphics{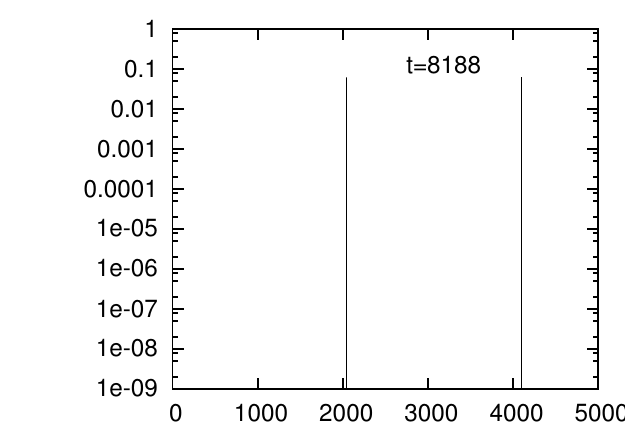}}
\scalebox{0.84}{\includegraphics{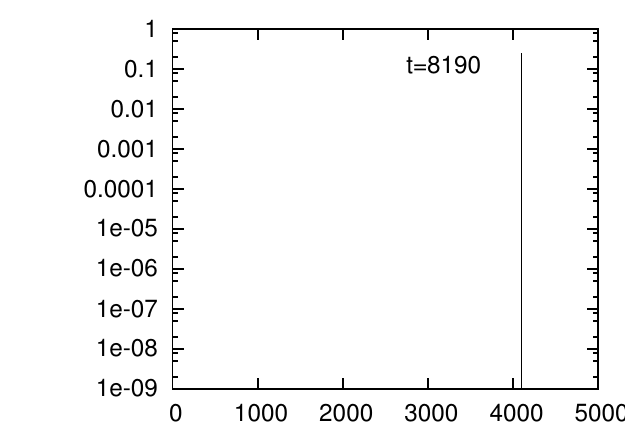}}
\end{center}
\caption{Power spectra of the configuration at 
t = 0, 4095, 8000, 8160, 8180, 8184, 8188, and 8190 (from top left to bottom right) in the evolution of rule 60 (vertical axis: $S(f)$, horizontal axis: $f$).}
\label{fig:r60fsp}
\end{figure}

\section{Power Spectral Analysis of configurations}

To investigate the periodicity of configurations emerging during the evolution of rule 60, we apply spectral analysis to the configurations. The discrete Fourier transform of a configuration $(a_0(t), a_1(t), \cdots a_{N-1}(t) )$ with array size $N$ at time step $t$ is given by
\begin{equation}
\hat{a}_t(f) = \frac{1}{N}\sum_{x=0}^{N-1}a_x(t){\exp}
(-i\frac{2{\pi}xf}{N}).
\label{eq:DFT}
\end{equation}

Let us define the power spectrum of CA as 
\[
S_t(f) = |\hat{a}_t(f)|^2.
\]

\begin{figure}
\begin{center}
\scalebox{0.84}{\includegraphics{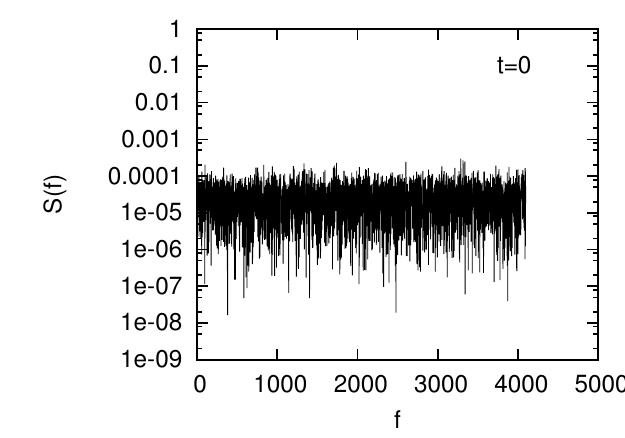}}
\scalebox{0.84}{\includegraphics{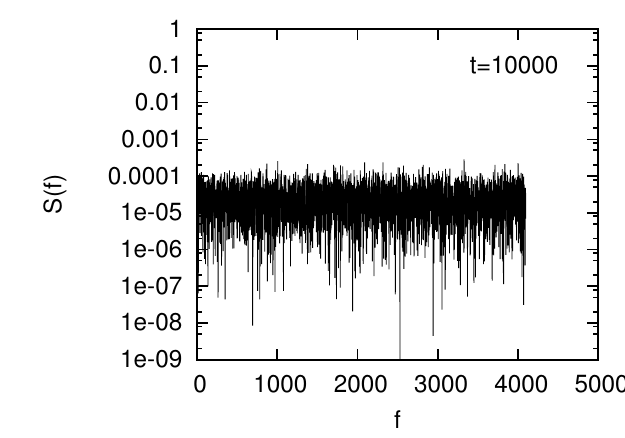}}
\end{center}
\caption{Power spectra of the configuration at time step $t=0$ (left) and $t=10000$ (right) in the evolution
of rule 60 starting from a random initial configuration with array size equal to 8193.}
\label{fig:r60N8193fsp}
\end{figure}

Figure \ref{fig:r60fsp} shows a typical example of the power spectra of the configuration at various time steps in the evolution of rule 60 starting from a random initial configuration of array size equal to $2^{13} =8192$ with odd parity. The power spectrum at $t$ = 0 is considered to be white noise because each site in the initial configuration takes state zero or state one randomly with independent equal probabilities. Although aperiodicity in the configuration continues to $t$ = 4094, periodicity suddenly emerges at $t$ = 4095. To be more specific, the components at even frequencies but zero in the spectrum become zero at $t$ = 4095. And the components at odd frequencies in the spectrum keep zero ever since $t$ = 4096. In general it is be proved that the components at even frequencies but zero in the power spectrum at $t=2^{n-1}-1$ is zero (Appendix D) and the ones at odd frequencies keep zero ever since $t = 2^{n-1}$ (Appendix E) in the CA rule defined in Eq.~(\ref{eq:Gentrns}) with initial configuration with odd parity of array size $2^n$. The power spectra become sparse after about $t$ = 8000 as the evolution proceeds. The component at the lowest frequency in the power spectrum at $t$ = 8160 is at $f$ = 256 that means the longest period in the configuration is 8192/256=32. The longest period in subsequent power spectrum is 16 at $t$ = 8180, 8 at $t$ = 8184, 4 at $t$ = 8188, and 2 at $t$ = 8190. This result implies that the periodicity contained in the configuration becomes short as the time step goes by and reminds us of the phenomenon in turbulence in viscous fluid called energy cascade process in which large scale eddies created by kinetic energy exerted by external force split into small scale ones by viscosity of fluid. Since every site has state one at time step $t$ = 8191, all the component except for $f$ = 0 in the spectrum is equal to zero.

As is well known that rule 60 generally exhibits chaotic behavior. Figure \ref{fig:r60N8193fsp} shows the power spectra obtained in the same way except for the array size equal to 8193. The power spectrum even at time step $t$=10000 exhibits white noise like a random initial configuration. In this case there is no regularity emerging during the evolution.

\section{Complexity of configurations}
Generally speaking, all the information necessary to perform a computational task by CA is included in its configuration. Therefore, it is reasonable to investigate the complexity of configuration during the process of solving the PP. As a measure of complexity, we employ Lempel-Ziv (LZ) complexity used in data compression called LZ78~\cite{ZL}. In LZ78, a string is divided into phrases. Given a string  $s_1s_2 \cdots s_ks_{k+1} \cdots$ where a substring $s_1s_2 \cdots s_k$ has already been divided into phrases $w_1 \cdots w_m \ (m \le k)$, the next phrase  $w_{m+1}$ is constructed by searching the longest substring $s_{k+1} \cdots s_{k+n} = w_j$, $(0 \le j \le m)$  and by setting $w_{m+1}=w_js_{k+n+1}$  where $w_0 = \epsilon$. The LZ complexity of the string is defined as the number of divided phrases. In this article we denote the LZ complexity of a given configuration at time step t by  $C_{LZ}(t)$.

\begin{figure}[t]
\begin{center}
\scalebox{0.7}{\includegraphics{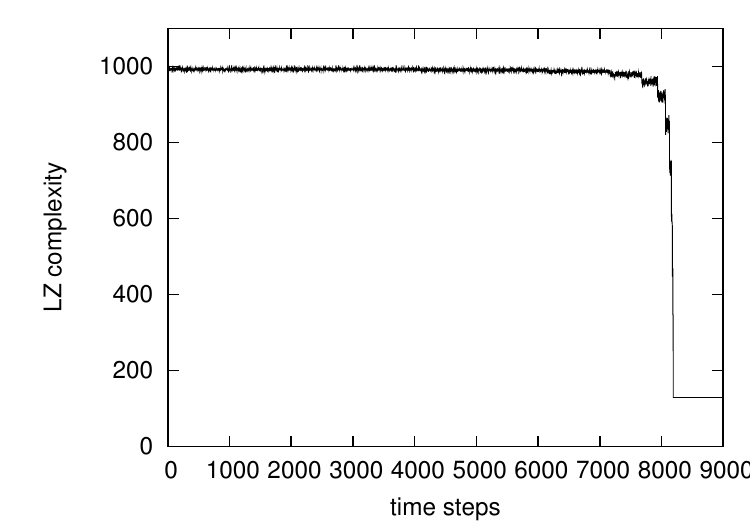}}
\scalebox{0.7}{\includegraphics{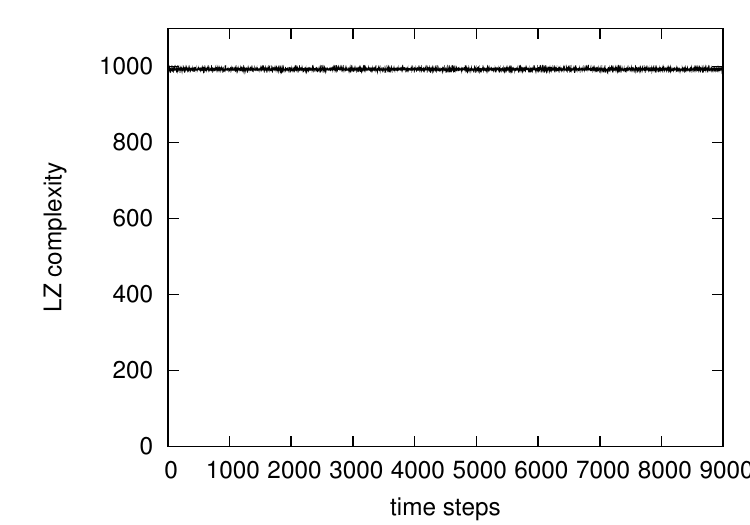}}
\end{center}
\caption{LZ complexity at each time step in the evolution of rule 60 starting from a random
initial configuration with array size equal to $2^{13}=8192$ (left) and 8193 (right).}
\label{fig:r60LZ}
\end{figure}

\begin{figure}
\begin{center}
\scalebox{0.7}{\includegraphics{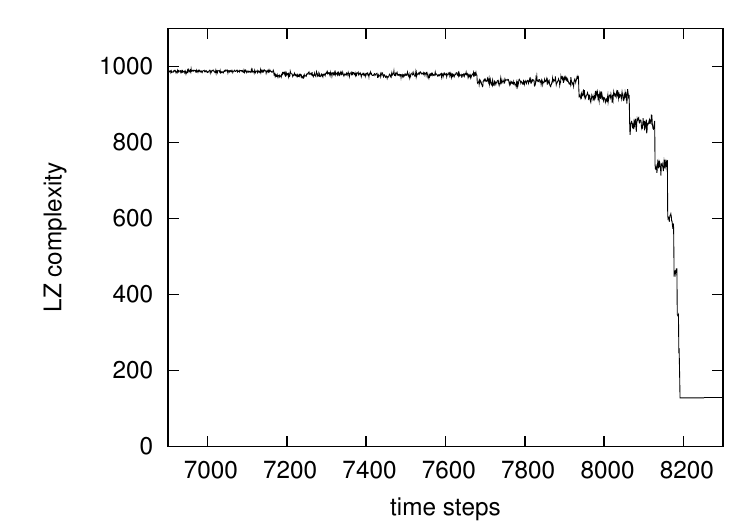}}
\scalebox{0.7}{\includegraphics{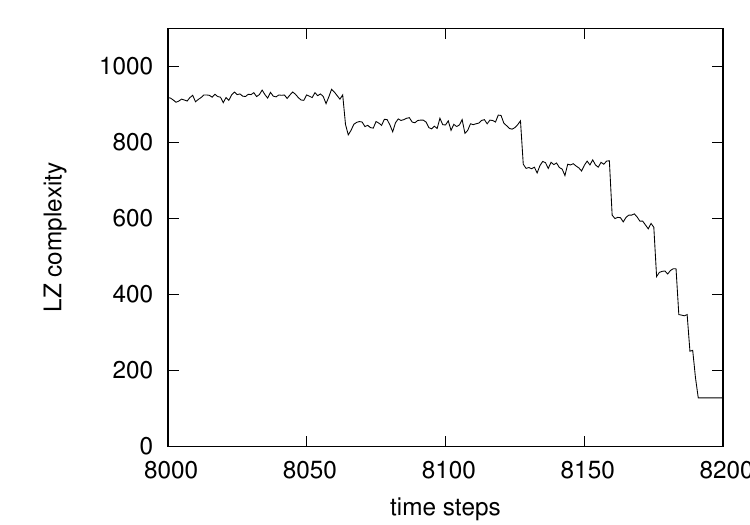}}
\end{center}
\caption{Enlarged view of the left of Fig. 5. There are stepping decreases in LZ complexity
at time step t=7168, 7680, 7936 on the left and at time step t=8064, 8128, 8160, 8176 on the right.}
\label{fig:r60LZ_fine}
\end{figure}

The left of Fig.~\ref{fig:r60LZ} shows the LZ complexity of configuration at each time step in the evolution of rule 60 starting from a random initial configuration with array size equal to $2^{13}$ = 8192 with odd parity. There is rapid decrease in  $C_{LZ}(t)$. after about $t$ = 7000. An enlarged view of the left of Fig.~\ref{fig:r60LZ} is Fig.~\ref{fig:r60LZ_fine}. On the left of Fig.~\ref{fig:r60LZ_fine} we can see stepping decreases in $C_{LZ}(t)$ at $t$ = 7168, $t$ = 7680, and $t$ = 7936 and on the right we can see the same at $t$ = 8064, $t$ = 8128, $t$ = 8160, and so on. The interval between stepping decreases in  $C_{LZ}(t)$ is the power of two and  $C_{LZ}(t)$ has roughly the same value in every single plateau. The average of LZ complexity $\overline{C_{LZ}(t)}$ in each plateau is described in detail in Table \ref{tbl:LZc}. The duration of the $j$-th plateau ($j \ge 0$)  seems to be $2^{n-j-1}$ in the size of array equal to  $2^n$  (n:positive integer). This stepping decrease of complexity is caused by the period halving in the evolution of additive rules. By letting $t$ = 0, $k=n-1$, $r$ = 1 in Eq.~(\ref{eq:E1}), we get
\[
a_i(2^{n-1}) = a_i(0) + a_{i-2^{n-1}}(0).
\]
This result means that the period of the configuration at time step $t = 2^{n-1}$ becomes $2^{n-1}$. Likewise the period becomes $2^{n-2}$ at $t = 2^{n-1} + 2^{n-2}$. These period halving processes recur n times until time step $t=2^n-1$. The period halving process exhibits a striking contrast to the persistent process of computation by the cellular automaton. In case of array size not equal to the power of two, there is no decrease in  $C_{LZ}(t)$ as shown on the right of Fig. \ref{fig:r60LZ}.

\begin{table}[t]
\caption{LZ complexity averaged on each plateau drawn on the left of Fig. 5.}
\begin{center}
\begin{tabular}{c|c|cp{1mm}c|c|c}
time steps  & duration & $\overline{C_{LZ}(t)}$ & & time steps  & duration & $\overline{C_{LZ}(t)}$ \\
\cline{1-3}\cline{5-7}
0 - 4095    & $2^{12}$  & 992.4 & & 8128 - 8159 & $2^{5}$   & 739.3 \\
4096 - 6143 & $2^{11}$  & 990.8 & & 8160 - 8175 & $2^{4}$   & 596.8 \\
6144 - 7167 & $2^{10}$  & 987.1 & & 8176 - 8183 & $2^{3}$   & 460.1 \\
7168 - 7679 & $2^{9}$   & 978.6 & & 8184 - 8187 & $2^{2}$   & 346.0 \\
7680 - 7935 & $2^{8}$   & 960.3 & & 8188 - 8189 & $2^{1}$   & 252.0 \\
7936 - 8063 & $2^{7}$   & 922.0 & & 8190 - 8190 & $2^{0}$   & 181.0 \\
8064 - 8127 & $2^{6}$   & 849.3 & & 8191 - $\infty$     & -  & 128.0
\end{tabular}
\label{tbl:LZc}
\end{center}
\end{table}

The PP is considered to be a decision problem in which the answer to an given instance is either yes or no. That means the answer is represented by one bit while the instance is encoded by $n$ bits ($n > 1$) in a general setting.  Therefore solving a decision problem is necessarily accompanied by the decrease in the complexity of information as the process of computation proceeds. The decrease in LZ complexity observed on the left of Fig. 5 implies that the evolution of rule 60 is a decision problem solving process, while almost the same amount of complexity during the evolution on the right of Fig.~\ref{fig:r60LZ} implies that the computational process of CA is not regarded as solving any decision problem.

\begin{figure}
\begin{center}
\scalebox{0.3}{\includegraphics{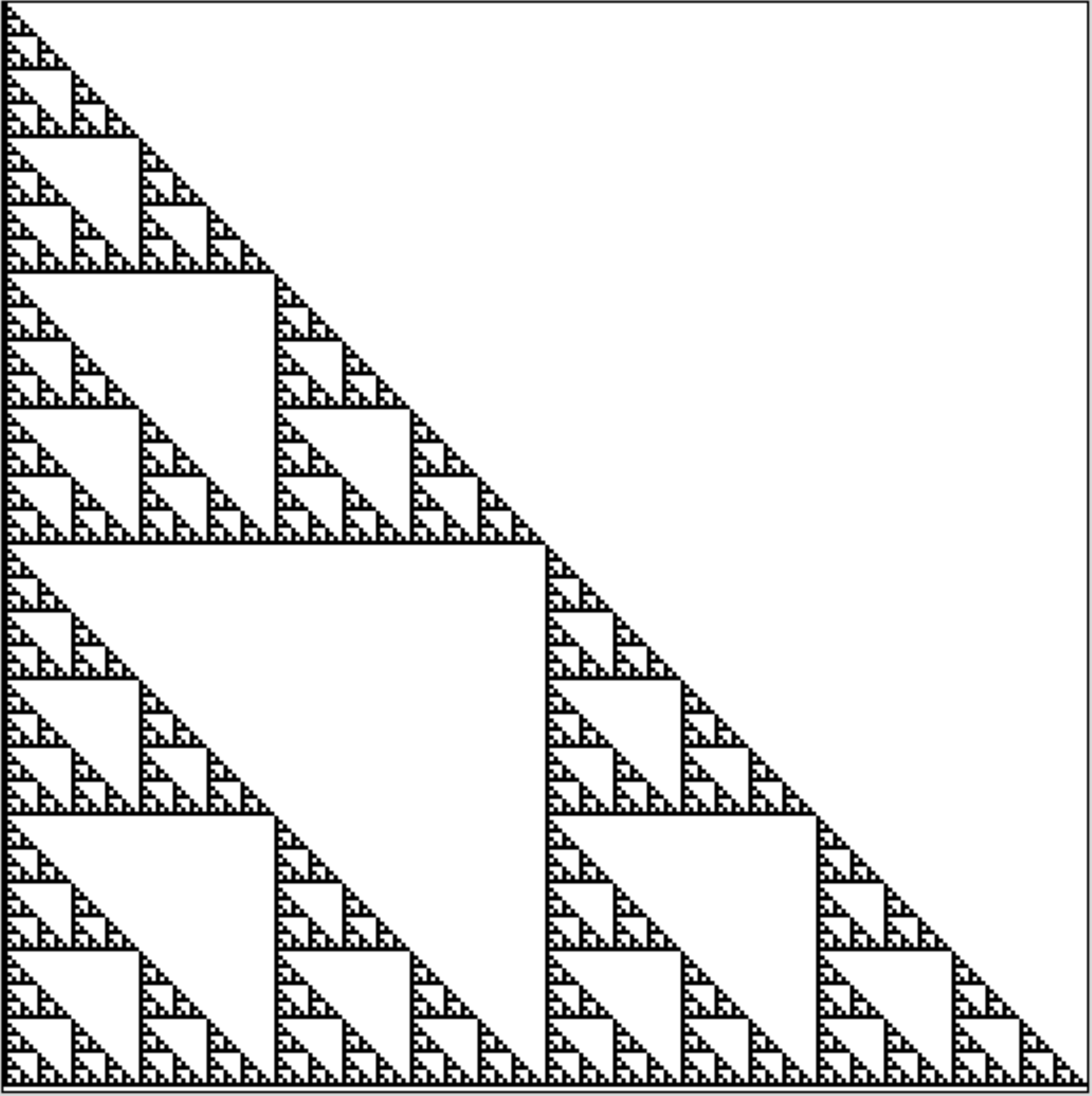}}
\end{center}
\caption{Space-time pattern of rule60 starting from a initial configuration with array size $256$
in which only the leftmost cell has state one. Every cell at $t=255$ has state one because the
parity of the initial configuration is odd.}
\label{fig:r60single}
\end{figure}

\begin{figure}
\begin{center}
\scalebox{0.7}{\includegraphics{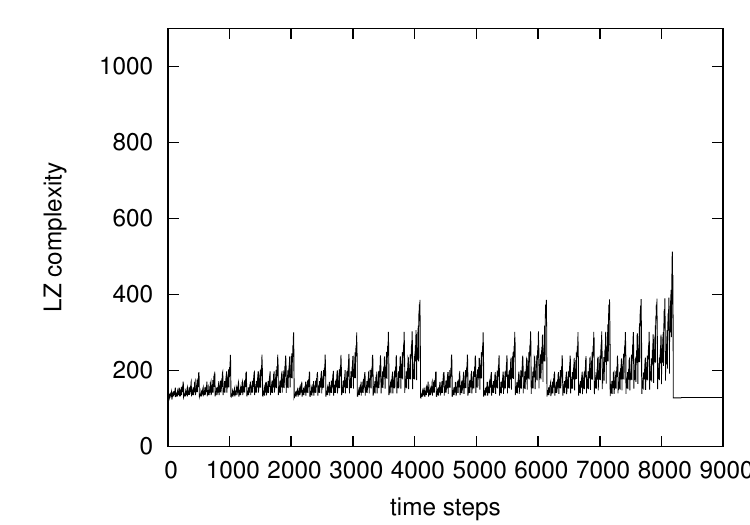}}
\scalebox{0.7}{\includegraphics{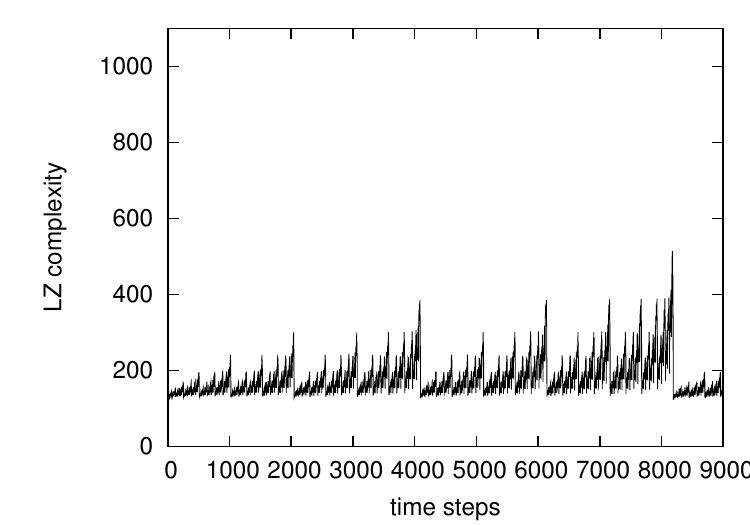}}
\end{center}
\caption{LZ complexity at each time step in the evolution of rule 60 starting from a initial
configuration with array size $2^{13}=8192$ (left) and $8193$ (right) in which  only the leftmost cell has state one.}
\label{fig:r60LZ_single}
\end{figure}

Next let us investigate the case in which the complexity involved in an initial configuration is small. Figure \ref{fig:r60single}  shows the space-time pattern of rule 60 starting from an initial configuration with array size 256 in which only the leftmost cell has state one. Every cell at $t$ = 255 has state one because the parity of the initial configuration is odd. Figure \ref{fig:r60LZ_single} shows LZ complexity at each time step in the evolution of rule 60 starting from an initial configuration with array size $2^{13}$ = 8192 (left) and 8193  (right) in which only the leftmost cell has state one. In the case of array size 8192,  $C_{LZ}(t)$  ($t \ge 8191$) keeps 128,  while $C_{LZ}(0)$ is 129. In this case, LZ complexity temporarily increases during the computation. This result is considered that the array is used as a working memory to hold temporary data necessary to obtain the answer. In case of array size 8193, $C_{LZ}(t=8191)$ is 128 and $C_{LZ}(t)$ continues fluctuating ever afterward.

\section{Conclusion}
We have studied the process of solving the PP by elementary CA rule 60 with array size $2^n$ ($n$: positive integer) by means of spectral analysis and LZ complexity. We observed the shift from longer period to shorter one in the power spectrum as the evolution proceeds that is reminiscent of cascade process in turbulence. Next we  observed the stepping decrease in LZ complexity every $2^k$ time steps ($k = n-1, n-2, \cdots,2,1,0$) due to the period halving process. These two phenomena, cascade process and period halving process, are observed in other CA rules defined by Eq.~(\ref{eq:Gentrns}).

Since the decline of LZ complexity during the evolution is not observed in the case of array size not equal to the power of two, this phenomenon might be peculiar to decision problem solving process. If it is true, we might be able to find decision problem solving CA by searching for the decline of LZ complexity during the evolution. In the future plan, we are going to study the possibility of using the LZ complexity to find CA rules that is able to solve a decision problem.

This paper is extended from the exploratory paper presented in AUTOMATA and JAC 2012~\cite{Nina12}.

\appendix 
\section{Additivity}
The configuration $( a_0(t), a_1(t), a_2(t), \cdots, a_{N-1}(t) )$ of the CA at time step $t$ is represented by the characteristic polynomial
\[
C(t) = \sum^{N-1}_{i=0}a_i(t)x^i,
\]
where $a_i(t)$ are integers modulo 2. The index $i$ in $a_i$ and $x_i$ is taken to be modulo $N$. Although \cite{Martin} employs dipolynomials in which positive and negative powers of $x$ appear, we use conventional polynomials with positive  powers of $x$ for its simplicity. We consider a ($r$+1)-neighbor CA whose transition rule is defined by Eq.~(\ref{eq:Gentrns}). The transition of characteristic polynomial $C(t)$ of the rule is represented by multiplication of polynomial $T(x) = (1+x^r)$ as follows
\[
C(t+1) = T(x) C(t) = \sum_{i=0}^{N-1}(a_i(t)+ a_{i-r}(t))x^i.
\]
To prove the additivity of the CA rule defined in Eq.~(\ref{eq:Gentrns}),  we will show that the both sides of the Eq.~(\ref{eq:add}) are represented by a same characteristic polynomial.
Consider any two configurations $\rho$ and $\tau$ represented by the characteristic polynomials $C_{\rho}$ and $C_{\tau}$ respectively.
\[
C_{\rho} =  \sum^{N-1}_{i=0}a_ix^i, \ \ C_{\tau} =  \sum^{N-1}_{i=0}b_ix^i.
\]
$\phi[\rho]$ is represented as follows.
\[
C_{\phi[\rho]} = (1+x^r)  \sum^{N-1}_{i=0}a_ix^i =  \sum^{N-1}_{i=0}(a_i + a_{i-r} )x^i.
\]
$\phi[\rho]$ is done in the same way.
Since the configuration $\rho \oplus \tau$  is represented by
\[
 \sum^{N-1}_{i=0}( a_i + b_i )x^i,
\]
the configuration $\phi[\rho \oplus \tau]$ is represented by
\[
C_{\phi[\rho \oplus  \tau]} = (1+x^r) \sum^{N-1}_{i=0}( a_i + b_i )x^i = \sum^{N-1}_{i=0}(a_i + a_{i-r} + b_i + b_{i-r})x^i = C_{\phi[\rho]} + C_{\phi[\tau]}
\]
Therefore the transition rule defined by Eq.~(\ref{eq:Gentrns}) satisfies the additivity.

\section{Solving the PP by additive CA in array size of the power of two.}
We consider characteristic polynomial at time step $t$.
\begin{equation}
C(t) = (1+x^r)^t C(0) = \sum_{j=0}^{t} {t  \choose  j} x^{jr} C(0).
\label{eq:B1}
\end{equation}
We consider the cases $r$ odd and $r$ even separately.

(a)  $r$ odd. we set $t = 2^n - 1$, ($n \ge 1$) in Eq.~(\ref{eq:B1}). We obtain
\[
C(2^n-1) = \sum^{N-1}_{i=0}\sum^{2^n-1}_{j=0}a_{i-jr}(0)x^i.
\]
By comparing the both sides of the equation, we have
\begin{equation}
a_i(2^n-1) = \sum^{2^n-1}_{j=0}a_{i-jr}(0).
\label{eq:B2}
\end{equation}
By letting $N = 2^n$ in Eq.~(\ref{eq:B2}), we obtain
\[
a_i(2^n-1) = \sum^{2^n-1}_{j=0}a_j(0),
\]
because gcd($2^n$, $r$) = 1  (sec. 4.8 in \cite{Knuth}). Therefore, the state of any cell at $t=2^n-1$ is equal to the parity of the initial configuration.

(b)  $r$ even. we define $k$ as
\[
k= \max \{ j \in \{0,1,2,\cdots \} | 2^j | r  \}
\]
where $a | b$ means that $b$ is divisible by $a$. By setting $t=2^{n-k}-1$, ($n > k$) in Eq.~(\ref{eq:B1}), we get
\[
C(2^{n-k}-1) = \sum^{N-1}_{i=0}\sum^{2^{n-k}-1}_{j=0}a_{i-jr}(0)x^i.
\]
Hence, we have
\begin{equation}
a_i(2^{n-k}-1) = \sum^{2^{n-k}-1}_{j=0}a_{i-jr}(0).
\label{eq:B3}
\end{equation}
In the case of $N = 2^n$, the  $2^{n-k}$  positions taken in the summation of Eq.~(\ref{eq:B3}) are different each other.
Therefore  the sum of any one of the consecutive $2^k$ cells at $t=2^{n-k}-1$  equals the parity of the initial configuration.

\section{Configurations with odd parity occur only as a initial configuration in additive CA}
We consider a configuration ¦Ò in which $i_1, i_2,\cdots$ and $i_n$-th cells have state one and the other cells have state zero. The configuration $\sigma$ is represented by the superposition of ${\sigma}_1, {\sigma}_2, \cdots, {\sigma}_n$, as follows.
\[
\sigma = \sigma_1 \oplus \sigma_2 \oplus \cdots \oplus \sigma_n
\]
where the configuration ${\sigma}_i$ has a state one cell only at site $i$. The configuration $\sigma$ yields
\begin{equation}
\phi[\sigma] = \phi[ {\sigma}_1 \oplus {\sigma}_2 \oplus \cdots \oplus {\sigma}_n ] =  \phi[{\sigma}_1] \oplus  \phi[{\sigma}_2] \oplus \cdots \oplus  \phi[{\sigma}_n]
\label{eq:C1}
\end{equation}
after one time step evolution. Since the configuration $\phi[{\sigma}_i]$ has state one cell only at distinct sites $i_k$ and $i_k+r$ ( we suppose $r$ is not divisible by $N$), we get $P(\phi[{\sigma}_i])=0$. Let denote $\sharp(\rho)$ the number of state one cells in configuration  $\rho$. It is clear that
\[
\sharp(\rho \oplus \tau) = \sharp(\rho) + \sharp(\tau) - 2\sharp(\rho  \wedge \tau),
\]
where $\wedge$ is logical AND operator. Therefore, for any configurations $\rho$ and $\tau$ with $P(\rho) = P(\tau) =0$, we get $P(\rho \oplus \tau)=0$. By Eq.~(\ref{eq:C1}) we get $P(\phi[\sigma]) = 0$.

\section{$S_{2^{n-1}-1}(2m) = 0, (m = 1, 2,\cdots)$ in additive CA with
 $r$ odd in case of odd initial parity in array size of the power of two.}
By setting $N = 2^n$ and $t = 2^{n-1}-1$ in Eq.~(\ref{eq:DFT}), we get 
\begin{eqnarray}
\hat{a}_{2^{n-1}-1}(f)  &=&
 \frac{1}{2^n}\sum_{x=0}^{2^n-1}a_x(2^{n-1}-1){\exp}(-i\frac{{\pi}xf}{2^{n-1}})
  \nonumber  \\
                                       &=&
				       \frac{1}{2^n}\sum_{x=0}^{2^{n-1}-1}[a_x(2^{n-1}-1)+(-1)^fa_{x+2^{n-1}}(2^{n-1}-1)]
				       \nonumber \\
			& &	       \times {\exp}(-i\frac{{\pi}xf}{2^{n-1}}) 
\label{eq:D1}
\end{eqnarray}
Since ${x \choose j}$ is odd for all $0 \leq j \leq x$ if and only if $x = 2^k -1, (k=0,1,2,\cdots)$ (exercise 6 in chapter 1 in \cite{Stanley}), it is not difficult to prove the following formula in the CA rule defined in Eq.~(\ref{eq:Gentrns}).
\begin{equation}
a_x(t+2^k-1) = \sum_{j=0}^{2^k-1}a_{x-rj}(t),  \ \ (k=0,1,2,\cdots).
\label{eq:D2}
\end{equation}
This formula holds true for any array size.
We obtain the following two formulas
\begin{eqnarray}
a_x(2^{n-1}-1) &=& \sum_{j=0}^{2^{n-1}-1}a_{x-rj}(0),  \nonumber \\
a_{x+2^{n-1}}(2^{n-1}-1) &=& \sum_{j=0}^{2^{n-1}-1}a_{x+2^{n-1}-rj}(0), \nonumber
\end{eqnarray}
from Eq.~(\ref{eq:D2}). Two indices $x-rj$ and $x+2^{n-1}-rj$ do not take the same vale for any $x$ and $j$ in the case of odd $r$. Since the parity of initial configuration is odd, the possible pair of ( $a_x(2^{n-1}-1), a_{x+2^{n-1}}(2^{n-1}-1)$ ) is (1, 0) or (0, 1). If $f$ is even, we get

\begin{equation}
\hat{a}_{2^{n-1}-1}(f) = 
\frac{1}{2^n}\sum_{x=0}^{2^{n-1}-1}{\exp}(-i\frac{{\pi}xf}{2^{n-1}}) = 0,  f \ne 0
\end{equation}
from Eq.~(\ref{eq:D1}).

\section{$S_t(2m+1)=0, (t \ge 2^{n-1}, m=0,1,2,\cdots)$ in additive CA with $r$ odd in array size of the power of two}
Before addressing this statement, we prove the formula,
\begin{equation}
a_i(t+2^k) = a_{i}(t) + a_{i-2^kr}(t),  \ \ (k=0,1,2,\cdots).
\label{eq:E1}
\end{equation}
in the CA rule defind in Eq.~(\ref{eq:Gentrns}). We consider the characteristic polynomial at time step $t+2^k$, 
\begin{eqnarray}
C(t+2^k) & = & \sum_{j=0}^{2^k}{2^k \choose   j} x^{jr}C(t) \nonumber  \\
               & = & (1+x^{2^kr})C(t)                       \nonumber \\
               & = & \sum_{i=0}^{N-1}[a_i(t)+a_{i-2^kr}(t)]x^i
\end{eqnarray}
We get Eq.~(\ref{eq:E1}) by comparing each term in the both sides of this formula. This formula holds true for any array size.

By setting $t = 2^{n-1} + t'$ ($t' \ge 0$) and $f = 2m+1$ ($m = 0, 1, 2,\cdots$), $N = 2^n$ in Eq.~(\ref{eq:DFT}), we get
\begin{eqnarray}
\hat{a}_{2^{n-1}+t'}(2m+1) &=& \frac{1}{2^n}\sum_{x=0}^{2^{n-1}-1}
 [a_x(2^{n-1}+t') - a_{x+2^{n-1}}(2^{n-1}+t')]  \nonumber \\
                           & & \times {\exp}(-i\frac{{\pi}(2m+1)x}{2^{n-1}}).
\label{eq:E2}
\end{eqnarray}
The expression in the summation yields
\begin{eqnarray}
\lefteqn{a_x(2^{n-1}+t') - a_{x+2^{n-1}}(2^{n-1}+t') } \nonumber \\
&=&                  a_x(t') + a_{x-2^{n-1}r}(t') - [ a_{x+2^{n-1}}(t') + a_{x+2^{n-1}-2^{n-1}r}(t') ]
\label{eq:E3}
\end{eqnarray}
by virtue of Eq.~(\ref{eq:E1}). Since $x$ is equal to $x + 2^{n-1} - 2^{n-1}r$ modulo $2^n$ and $x -2^{n-1}r$ is equal to $x + 2^{n-1}$ modulo $2^n$ in the case of $r$ odd, 
the left hand side of Eq.~(\ref{eq:E3}) is equal to zero. So Eq.~(\ref{eq:E2}) becomes zero.

\end{document}